\documentclass[a4paper]{jpconf}
\usepackage{graphicx}
\graphicspath{{images/}}
\usepackage{amsmath}
\usepackage{algorithm}
\usepackage{algorithmic}

\begin{document}
\title{Efficient Multi-site Data Movement Using Constraint Programming for Data Hungry Science}

\author{Michal Zerola}
\address{Nuclear Physics Institute, Academy of Sciences of the Czech
  Republic, Czech Republic}
\ead{michal.zerola@ujf.cas.cz}

\author{J\'{e}r\^{o}me Lauret}
\address{Brookhaven National Laboratory, Upton, USA}
\ead{jlauret@bnl.gov}

\author{Roman Bart\'{a}k}
\address{Faculty of Mathematics and Physics, Charles University, Czech Republic}
\ead{roman.bartak@mff.cuni.cz}

\author{Michal \v{S}umbera}
\address{Nuclear Physics Institute, Academy of Sciences of the Czech
  Republic, Czech Republic}
\ead{sumbera@ujf.cas.cz}

\begin{abstract}
For the past decade, HENP experiments have been heading towards a
distributed computing model in an effort to concurrently process tasks
over enormous data sets that have been increasing in size as a
function of time. In order to optimize all available resources
(geographically spread) and minimize the processing time, it is
necessary to face also the question of efficient data transfers and
placements. A key question is whether the time penalty for moving the
data to the computational resources is worth the presumed gain. Onward
to the truly distributed task scheduling we present the technique
using a Constraint Programming (CP) approach. The CP technique
schedules data transfers from multiple resources considering all
available paths of diverse characteristic (capacity, sharing and
storage) having minimum user's waiting time as an objective. We
introduce a model for planning data transfers to a single destination
(data transfer) as well as its extension for an optimal data set
spreading strategy (data placement). Several enhancements for a solver
of the CP model will be shown, leading to a faster schedule
computation time using symmetry breaking, branch cutting, well studied
principles from job-shop scheduling field and several
heuristics. Finally, we will present the design and implementation of
a corner-stone application aimed at moving datasets according to the
schedule. Results will include comparison of performance and trade-off
between CP techniques and a Peer-2-Peer model from simulation
framework as well as the real case scenario taken from a practical
usage of a CP scheduler.
\end{abstract}

\section{Introduction}
Paramount to a distributed computing model (Grid or Cloud) is the key
problem of processing in the most efficient manner vast amount of data
in a minimum time. Since the beginning of the decade, High Energy and
Nuclear Physics (HENP) communities have been tackling this challenging
problem with an ultimate focus to optimize all of their available
resources (geographically spread) and thus minimize the processing
time it takes to go over their steadily growing data sets.

One of such communities is the computationally challenging experiment
STAR at RHIC (Relativistic Heavy Ion Collider \cite{adams-2005-757})
located at the Brookhaven National Laboratory (USA). In addition to a
typical Peta-byte scale storage requirements and large computational
need this experiment as a running experiment acquires a new set of
valuable experimental data every year, introducing other dimension of
safe data transfer to the problem. From the yearly data sets, the
experiment may produce many physics-ready derived data sets which
differ in accuracy as the problem is better understood and as time
passes.

The user's task is typically embarrassingly parallel; that is, a
single program can run $N$ times on a fraction of the whole data set
split into $N$ sub-parts with usually no impact on science
reliability, accuracy, or reproducibility. For a computer scientist,
the issue then becomes how to split the embarrassingly parallel task
into $N$ jobs in the most efficient manner while knowing the data set
is spread over the world and/or how to spread 'a' dataset and place
best the data for maximal efficiency and fastest processing of the
task.

Rather than trying to solve the whole complex issue including optimal
data placement strategy (distribution of centrally acquired data to
other processing sites) with an emphasis on efficient further
processing we split the problem into several stages. In this paper we
focus on one block of this complex task which is of immediate need by
the physicists: ``how to bring the desired dataset to a single
destination in the shortest time?'' By isolating the data
transfer/placement and the computational challenges from each other,
we get an opportunity to study the behavior of both sets of
constraints separately. The paper summarize the work from
\cite{zerola-icaps} addressed to more theoretically based audience
from automated planning community. In addition we present extensions
for other real-life requirements in \ref{subsec:add_const} and propose
an architecture for further implementation in \ref{subsec:arch}.

\section{Related works}
The needs of large-scale data intensive projects arising out of
several fields such as bio-informatics (BIRN, BLAST), astronomy (SDSS)
or HENP communities (STAR, ALICE) have been the brainteasers for
computer scientists for years. Whilst the cost of storage space
rapidly decreases and computational power allows scientists to analyze
more and more acquired data, appetite for efficiency in Data Grids
becomes even more of a prominent need.

Decoupling of job scheduling from data movement was studied by
Ranganathan and Foster in \cite{10.1109/HPDC.2002.1029935}. The
authors discussed combinations of replication strategies and
scheduling algorithms, but not considering the performance of the
network. The nature of high-energy physics experiments, where data are
centrally acquired, implies that replication to geographically spread
sites is a must in order to process data distributively. Intention to
access large-scale data remotely over wide-area network has turned out
to be highly ineffective and a cause of often poorly traceable
troubles.

The authors of \cite{1272383} proposed and implemented improvements to
Condor, a popular cluster-based distributed computing system. The
presented data management architecture is based on exploiting the
workflow and utilizing data dependencies between jobs through study of
related DAGs. Since the workflow in high-energy data analysis is
typically simple and embarrassingly parallel without dependencies
between jobs these techniques don't lead to a fundamental optimization
in this field.

Sato et al. in \cite{conf/grid/SatoMEM08} and authors of
\cite{conf/ccgrid/RahmanBA07a} tackled the question of replica
placement strategies via mathematical constraints modeling an
optimization problem in the Grid environment. The solving approach in
\cite{conf/grid/SatoMEM08} is based on integer linear programming
while \cite{conf/ccgrid/RahmanBA07a} uses a Lagrangian relaxation
method \cite{Fisher:lagrange}. The limitation of both models is a
characterization of data transfers which neglects possible transfer
paths and fetching data from a site in parallel via multiple links
possibly leading to better network utilization.

We focus on this missing component considering wide-area network data
transfers pursuing more efficient data movement. An initial idea of
our presented model originates from Simonis \cite{SimonisHCP} and the
proposed constraints for the traffic placement problem were expanded
primarily on link throughputs and consequently on follow-up transfer
allocations in time. The solving approach is based on the Constraint
Programming technique \cite{Marriott98:programming}, used in
artificial intelligence and operations research. One of the immense
advantages of the constrained based approach is a gentle augmentation
of the model with additional real-life rules.

\section{Formal model}
The input of the problem consists of two parts. The first part
represents the (Grid) network and file origins. The network, formally
a directed weighted graph, consists of a set of nodes $\mathbf N$
(sites) and a set of directed edges $\mathbf E$ (links). The weight of
an edge describes the number of time units needed to transfer a file
of one size unit. Information about files' origins is a mapping of
each file to a set of sites where the file is available. The second
part of the input is a user request, namely the set of files that need
to be transferred to a common destination site. The solving process is
composed of two stages:
\begin{itemize}
\item a transfer path for each file, i.e., one origin and a valid
  path from the origin to the destination, is selected ({\bf
    planning})
\item for each file and its selected transfer path, the particular
  transfers via links are scheduled in time such that the resulting
  plan has minimal makespan ({\bf scheduling})
\end{itemize}

The goal of the scheduling stage is to evaluate the path configuration
in the sense of the required makespan. Essentially, it works as an
objective function because the realization of the schedule will not
depend on particular transfer times calculated in this phase, as we
will show in section \ref{sec:sched_exec}.

Both stages iterate until the plan of transfers with the minimal
makespan is found (see Alg. \ref{code:search}). As we can see, the
phases are not strictly separated, while planning function takes {\it
  makespan} as an argument. This allows to prune the search space
already during generating transfer paths using constraint
\ref{eq:cut_const} explained latterly. In \cite{zerola-icaps} we
identified that about $90\%$ of overall time is spent in the planning
stage hence we put our effort to improve this stage.
\begin{algorithm}
  \caption{Pseudocode for a search procedure.}
  \label{code:search}
  \begin{algorithmic}
    \STATE makespan $\leftarrow \sup$
    \STATE plan $\leftarrow$ {\bf Planner}.getFirstPlan()
    \WHILE {plan != null}
    \STATE {schedule $\leftarrow$} {\bf Scheduler}.getSchedule(plan,
    makespan) \COMMENT{Branch-and-Bound on makespan}
    \IF{schedule.getMakespan() $<$ makespan}
    \STATE makespan $\leftarrow$ schedule.getMakespan()
    \COMMENT{better schedule found}
    \ENDIF
    \STATE {\bf Planner}.getNextPlan(makespan) \COMMENT{next feasible
      plan with cut constraint}
    \ENDWHILE
  \end{algorithmic}
\end{algorithm}
The following formalism is used to define a constraint model
describing the planning sub-problem. The set $\mathbf{OUT}(n)$
consists of all edges leaving node $n$, the set $\mathbf{IN}(n)$ of
all edges leading to node $n$. Input received from a user is a set of
demands $\mathbf D$ needed at the destination site $dest$. In our
notation demands represent file requests and we will use this
symbolism in the following text. For every demand $d\in D$ we have a
set of sources $\mathbf{orig}(d)$ - sites where the demanded file $d$
is already available. We will present the {\it link-based} approach
for modeling planning constraints. Another approach called {\it
  path-based} can be found in \cite{zerola-icaps}.

The essential idea of the link-based approach is using one decision
$\{0,1\}$ variable $X_{de}$ for each demand and link of the network,
denoting whether demand $d$ is routed over edge $e$ or
not. Constraints (\ref{eq:lcor}-\ref{eq:lcoth}), ensure that if all
decision variables have assigned values then the resulting
configuration contains transfer paths. These constraints alone allow
isolated loops along with the valid paths and therefore {\it
  precedence constraints} (\ref{eq:precc}) are used to eliminate such
loops.
\begin{equation}
  \label{eq:lcor}
  \forall d\in \mathbf D: \sum_{e\in \cup \mathbf{OUT}(n|n\in
    \mathbf{orig}(d))}X_{de} = 1, \, \sum_{e\in \cup
    \mathbf{IN}(n|n\in \mathbf{orig}(d))}X_{de} = 0
\end{equation}
    
\begin{equation}
  \label{eq:lcdest}
  \forall d\in \mathbf D: \sum_{e\in \mathbf{OUT}(dest(d))}X_{de} = 0,
  \; \sum_{e\in \mathbf{IN}(dest(d))}X_{de} = 1
\end{equation}

\begin{equation}
  \label{eq:lcoth}
  \begin{split}
  &\forall d\in \mathbf D, \forall n\notin orig(d)\cup
    \{dest(d)\}:\\
    \sum_{e\in \mathbf{OUT}(n)}X_{de} \leq 1,& \;
    \sum_{e\in \mathbf{IN}(n)}X_{de} \leq 1, \;\sum_{e\in
      \mathbf{OUT}(n)}X_{de} = \sum_{e\in \mathbf{IN}(n)}X_{de}
  \end{split}
\end{equation}
Precedence constraints (\ref{eq:precc}) use non-decision positive
integer variables $P_{de}$ representing possible start times of
transfer for demand $d$ over edge $e$. Let $dur_{de}$ be the constant
duration of transfer of $d$ over edge $e$. Then constraint
\begin{equation}
  \label{eq:precc}
  \forall d\in \mathbf{D}\; \forall n\in \mathbf{N}: \sum_{e\in \mathbf{IN}(n)}X_{de}\cdot (P_{de}+dur_{de})\leq
  \sum_{e\in \mathbf{OUT}(n)}X_{de}\cdot P_{de}
\end{equation}
ensures a correct order between transfers for every demand, thus
restricting loops. Unfortunately, constraints (\ref{eq:precc}) do not
restrict the domains of $P_{de}$ until the values $X_{de}$ are known
and therefore we suggest using a redundant constraint (\ref{eq:fil})
to estimate better the lower bound for each $P_{de}$. Let $start$ be
the start vertex of $e$ not containing demand $d$ ($start\notin
orig(d)$):
\begin{equation}
  \label{eq:fil}
  \min_{f\in \mathbf{IN}(start)}(P_{df}+dur_{df})\leq P_{de}
\end{equation}
Variables $P_{de}$ can be used not only to break cycles but also to
estimate the makespan of the plan. The idea is that according to the
number of currently assigned demands per some link and their possible
starting times, we can determine the lower bound of the makespan for
the schedule that will be computed later in the scheduling
stage. Hence if we have some upper bound for the makespan (typically
obtained as the best solution from the previous iteration of planning
and scheduling) we can restrict plans in next iterations by the
following constraint:
\begin{equation}
  \label{eq:cut_const}
  \forall e\in \mathbf{E}: \min_{d\in \mathbf{D}}(P_{de}) + \sum_{d\in
    \mathbf D}X_{de}\cdot dur_{de} + SP_{e} < makespan,
\end{equation}
where $SP_e$ stands for the value of the shortest path from the ending
site of $e$ to $dest$.

\section{Search heuristics}
The constraint model needs to be accompanied by a clever branching
strategy to achieve good runtimes. A clever branching strategy is a
key ingredient of any constraint satisfaction approach, especially as
the problem is NP-hard (as we have proven in \cite{zerola-icaps}).

According to the measurements shown in \cite{zerola-icaps}, the
majority of time was spent in the planning phase, hence we proposed an
improved variable selection heuristic that exploits better the actual
transfer times by using information from variables $P_{de}$. In
particular, the heuristic, called {\it MinPath}, suggests to
instantiate first variable $X_{de}$ such that the following value is
minimal:
\begin{equation}
  \inf{P_{de}} + dur_{de} + SP_{e},
\end{equation}
where $\inf{P_{de}}$ means the smallest value in the current domain of
$P_{de}$.



Concerning the value selection heuristics, both variants were tested,
particularly {\it Increasing} (assign first $0$, then $1$) and {\it
  Decreasing} (assign first $1$, then $0$) value iteration order.

In the scheduling phase two approaches were considered. First one,
called {\it SetTimes}, is based on determining Pareto-optimal
trade-offs between makespan and resource peak capacity. A detailed
description and explanation of this approach can be found in
{\cite{pape.couro-1994:time-compr:inbook:}}. The second one is a
texture-based heuristic called {\it SumHeight}, using ordering tasks
on unary resources. The implementation used originates from
{\cite{DBLP:conf/aaai/BeckDSF97a}} and supports {\it Centroid}
sequencing of the most critical activities.

\subsection{Additional real-life constraints}
\label{subsec:add_const}
The constraint model, formally presented in the previous sections,
covers mostly fundamental and essential attributes required in order
to solve efficient data transfers. However, reality can provide us
with further restrictions we have to deal with, as far as we intend to
tighten the gap between simulation and real life production.

\subsubsection{Storage space capacity.}
One of the facts we have suppressed during scheduling file transfers
via selected paths is a storage space limitation at sites. If a file
is going to be transferred through a site, there must be guaranteed
that from the start of a transfer over incoming link till the end of a
transfer over the outgoing link, enough storage space is available at
the relevant site. In fact, due to the current disk price, there will
hardly be the limitations of holding the bulk of files at intermediate
sites for a quite short amount of time, but the model is capable to
deal with it as well.

This restriction can be easily achieved by introducing a {\it
  cumulative} resource for each site with the capacity equal to the
free space of the site. If the demand $d$ enters the site via the link
$\mathit{inc}$ and leaves it via the link $\mathit{out}$ we create a
new task assigned to this cumulative resource. The start time of the
task will be equal to the start time of the task
$\mathbf{T}_{d,\mathrm{inc}}$ while the finish time to the finish time
of the task $\mathbf{T}_{d,\mathrm{out}}$, as depicted in Fig.
{\ref{fig:storage_space_resource}.

\begin{figure}
  \begin{center}
    \includegraphics[width=4in]{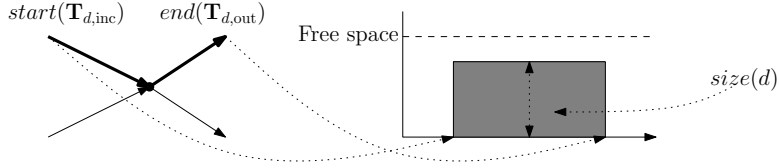}
    \caption{For a demand $d$ passing a site with limited space (via
      the links $\mathit{inc}$ and $\mathit{out}$) a new task is
      created with the starting and ending times set according to file
      transfers $\mathbf{T}_{d,\mathrm{inc}}$ and
      $\mathbf{T}_{d,\mathrm{out}}$.}
    \label{fig:storage_space_resource}
  \end{center}
\end{figure}

\subsubsection{Shared links.}
So far, we have assumed that all links incoming or outgoing from any
site have their own bandwidth (slowdown factor) that is not affected
by others. Nevertheless, in reality this is not always feasible, since
several links leading to a site usually share the same router and/or
physical fiber having throughput less than the sum of their own
values. Hence, one can't use such links simultaneously at their
maximum bandwidths.

We express this restriction by adding {\it dummy} vertices and edges
to the network graph and by modifying relevant slowdown factors. For a
site where the restriction exists due to the shared fiber or router, a
new dummy vertex is added together with a dummy edge, a connection
with the original site vertex. The slowdown factor of the added edge
is set to the real limitation and slowdown factors of affected shared
links are reduced by this amount.

\begin{figure}
  \begin{center}
    \includegraphics[width=4in]{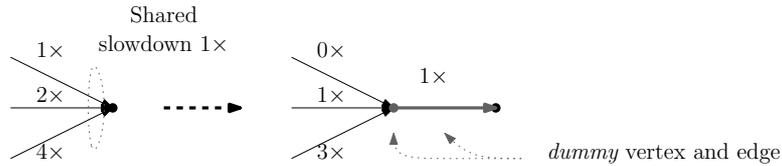}
    \caption{An illustration of a dummy vertex and edge insertion into
      the graph. Original slowdown factors of links affected are
      reduced by $1$, corresponding in this case to a limitation of a
      sharing resource.}
    \label{fig:shared_bandwidth}
  \end{center}
\end{figure}

\section{Comparative studies}
\label{sec:comp_stud}
We have implemented and compared performance of alternatives of the
model, namely using {\it link-based} and {\it path-based}
approaches. Several combinations of heuristics were tried and in
addition a comparison with a simulated Peer-2-Peer method is shown.

For implementation of the solver we use {\bf
  Choco} \footnote{http://choco.sourceforge.net}, a Java based library
for constraint programming. The Java based platform allows us an
easier integration with already existing tools in the STAR
environment.

\subsection{Peer-2-Peer (P2P) simulator}
The P2P model is well known and successfully used in areas such as
file sharing, telecommunication or media streaming. P2P model doesn't
allow file transfers via paths, only by direct connections. We
implemented a P2P simulator by creating the following work-flow: {\bf
  a)} put an observer for each link leading from an origin to the
destination; {\bf b)} if an observer detects the link is free, it
picks up the file at his site (link starting node), initiates the
transfer, and waits until the transfer is done. We introduced a
heuristic for picking up a file as typically done for P2P. The link
observer picks up a file that is
available at the smallest number of sites. If there are more files
available with the same cardinality of $orig(n)$, it randomly picks
any of them. After each transfer, the file record is removed from the
list of possibilities over all sites. This process is typically
resolved using distributed hash table (DHT)
\cite{DBLP:conf/iptps/NaorW03}, however in our simulator only simple
structures were used. Finally an algorithm terminates when all files
reach the destination, thus no observer has any more work to do.

\subsection{Data sets}
Regarding the data input part, the realistic-like network graph
consists of $5$ sites, denoted as BNL, LBNL, MIT, KISTI, and Prague
and all requested files are supposed to be transferred to the Prague
node.  The distribution of file origins, i.e. amount of files
available at one particular site, is following: $100\%$ of files are
available at a central repository at BNL, LBNL holds $60\%$, MIT
$1\%$, and KISTI $5\%$ of all files. Implemented demands feeder
generates the requested number of demands and for each demand decides
with a probability whether it is available at a particular site or
not, respecting the given distribution.

\subsection{Experiments}
Our experiments are designed to focus on evaluation of proposed
alternatives of the model and detecting the most suitable combination
of search heuristics. Understanding the performance and limitation of
the studied techniques in a simulated realistic environment is a
necessary step prior to further software deployment. All presented
experiments were performed on Intel Core2 Duo CPU@1.6GHz with 2GB of
RAM, running a Debian GNU Linux operating system.

We compared the performance of the {\it FastestLink} and {\it MinPath}
heuristics and a Peer-2-Peer model \cite{DBLP:conf/iptps/2003} that is
currently the most frequently used approach to solve the problem.

Figure
\ref{fig:heuristics_perf} shows that convergence of the new {\it
  MinPath} heuristic is faster than the {\it FastestLink} and both
heuristics achieve better makespan than the P2P approach.
\begin{figure}
  \begin{center}
    \includegraphics[width=3in]{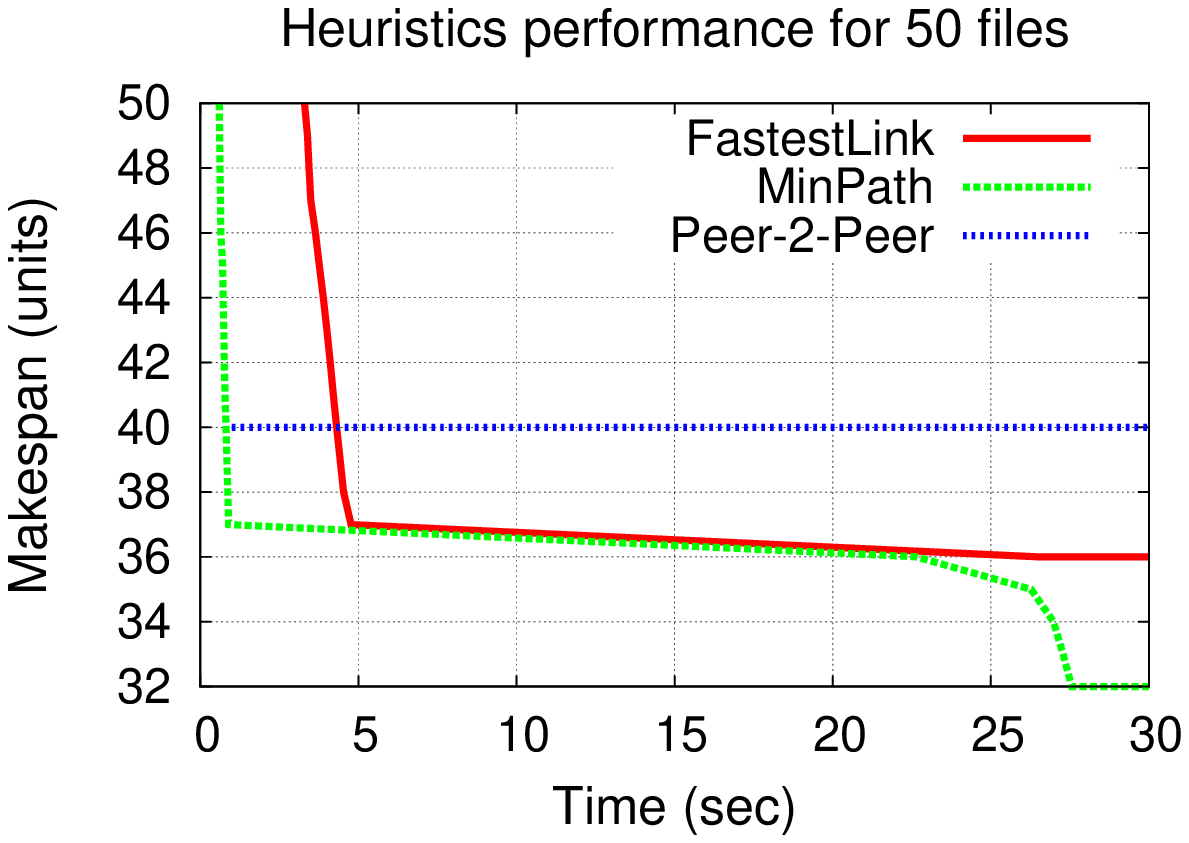}
    \includegraphics[width=3in]{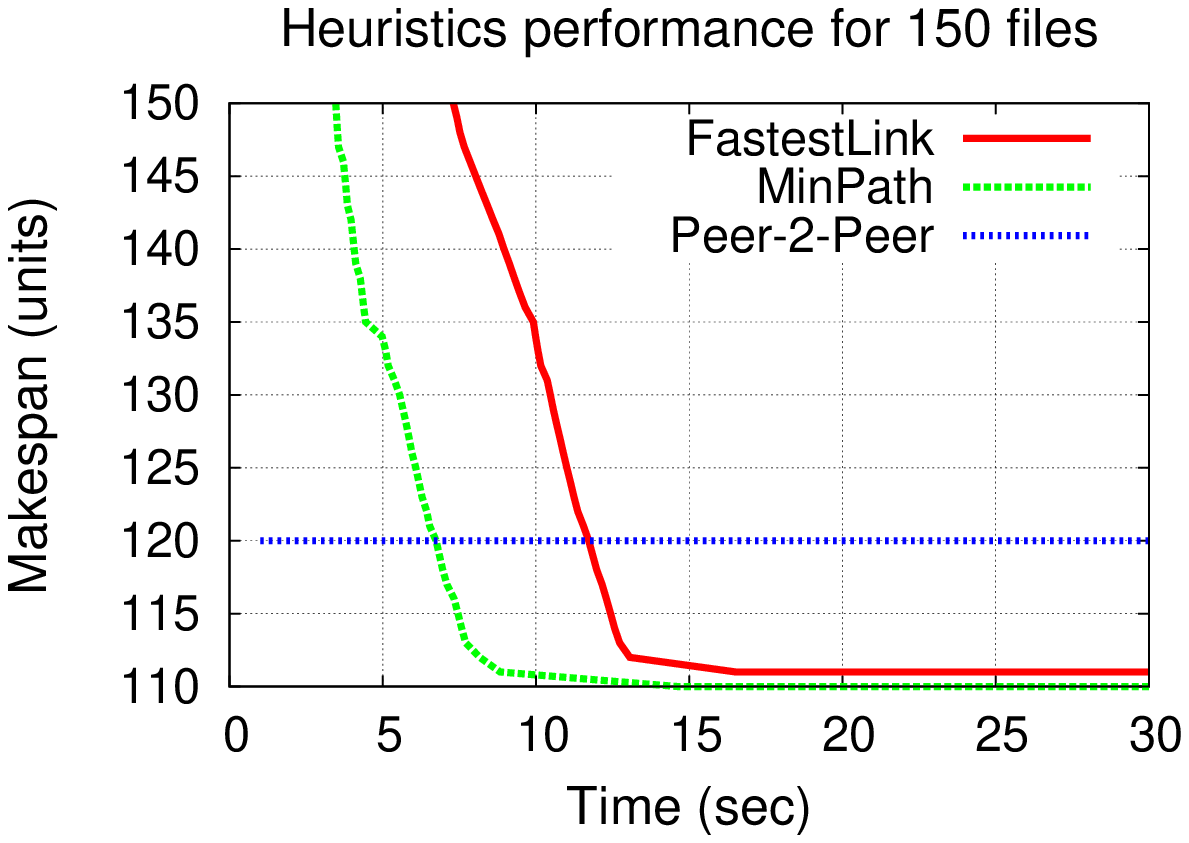}
    \caption{Convergence of makespan during the search process for
      {\it FastestLink} and {\it MinPath}.}
    \label{fig:heuristics_perf}
  \end{center}
\end{figure}
Table \ref{tab:makespan_cmp} shows similar comparison of heuristics
and the P2P model including the time when the best solution was found
for several input instances.
\begin{table}
  \begin{center}
    \begin{tabular}{c|c|c|c|c|c|}
      \cline{2-6} & \multicolumn{2}{c}{{\bf Solution time}} & \multicolumn{3}{|c|}{{\bf Makespan}}\\ \hline
      \multicolumn{1}{|c||}{\bf Files} & {\it FastestLink} & {\it MinPath} & {\it FastestLink} & {\it MinPath} & {\it P2P}\\ \hline\hline
      \multicolumn{1}{|c||}{\bf 25} & 3.862 & 1.431 & 14 & 14 & 24\\
      \multicolumn{1}{|c||}{\bf 50} & 26.508 & 27.556 & 36 & 32 & 40\\
      \multicolumn{1}{|c||}{\bf 100} & 8.627 & 3.176 & 73 & 73 & 80\\
      \multicolumn{1}{|c||}{\bf 150} & 16.52 & 14.618 & 111 & 110 & 120\\
      \multicolumn{1}{|c||}{\bf 200} & 26.167 & 14.031 & 146 & 146 & 160\\ \hline
    \end{tabular}
    \caption{Comparison of heuristics with emphasis on time when the best
      solution was found and the makespan.}
    \label{tab:makespan_cmp}
  \end{center}
\end{table}

As stated above, in reality the network characteristic is dynamic and
fluctuates in time. Hence, trying to create a plan for $~1000$ or more
files that will take several hours to execute is needless, as after
the time elapsed the computed plan does not have to be valid
anymore. Our intended approach is to work with batches of files, that
gives us another benefit of implementing fair-share mechanism in a
multi user environment as well. Particularly, the requests coming from
users are queued and differ in size and priorities of users. The
availability to pick demands from waiting requests into actual batch
within reasonably short intervals is very convenient for achieving
fair-shareness. The experiments give us an estimate on the number of
files per batch.

\section{Schedule execution}
\label{sec:sched_exec}
In this section we will briefly discuss the applicability of our model
and approach in a real-life network and protocol mechanism and explain
the suggested schedule execution in such an environment. In order to
achieve a full link speed between two points in a wide area network
(to fully saturate the bandwidth) one has to understand the basic
principles of the TCP/IP protocol communication. An operation of
transferring a single data file itself consists of splitting the data
content into packets of a given size (defined by {\it window scaling}
parameter of the protocol) and sending them one-by-one from the source
to the destination.

Since the reception of each packet has to be acknowledged by the
receiver to achieve both data integrity and delivery guarantee, the
time for the acknowledged packet to travel from source to destination
and back, so called {\it round-trip time (RTT)} plays an important
role. In a wide area network the {\it RTT} is usually significant, and
to overcome such delays, current data transfer tools use threads to
handle several TCP streams in parallel in an attempt to smooth or
minimize the intrinsic delays associated with TCP. Another standard
approach in HENP communities is the execution of several data
transfers in parallel (multiple sender nodes per link) to increase the
bandwidth usage. With this last approach, any one instance downtime
would be compensated by other active senders’ transfers. Both
approaches are typically combined for best results and it has been
experimentally shown and taken as a standard assumption that a flat
transfer rate could be achieved across long distance.

The presented model in previous sections assumes a single file
transfer at any time on a link using {\it unary} resources. Trying to
explicitly model the real network and packets behavior would hardly
lead to any optimization (following all network peculiarities would
cause the model to be barely realizable). The mechanism of how the
computed schedule will be executed in the real network is following:
\begin{itemize}
\item every link is supplied by one {\it LinkManager} that is
  responsible for transferring files over the link if the link is part
  of their computed transfer path
\item as soon as a file becomes available, the corresponding {\it
  LinkManager} initiates another instance of the transfer, respecting
  a maximum allowed simultaneous parallel instances
\end{itemize}
So rather than following the exact schedule, the implementation
considers just the plan - the computed transfer paths, because there
is no {\it due-time} limitation and executes transfers in a {\it
  greedy} manner. However, to allow this, one has to be sure that the
computed time to complete the schedule would not differ substantially
from the real execution of the transfers. Consequently we developed
the realistic network simulator along the above facts and comparison
of the makespans showed results consistent with each other within a
3\% margin, which is negligible. Hence the experiment confirmed that
presented model provides a fairly accurate estimate of the real
makespan.

\subsection{Architecture}
\label{subsec:arch}
We will briefly sketch out the proposed architecture of the automated
data transfer planning and executing. The requests from the users are
collected and recorded in a relational database. One convenient
approach is to use a web interface backed by a framework such as {\bf
  Django} \footnote{http://www.djangoproject.com} that handles forms
and templates and offers plugins to several common relational
databases. The planner, a standalone central component (the brain of
the system) selects the batch of requested files from the database and
computes the plan for it. The selection process depends on the
fair-share objectivity function and allows us to modularly implement
and test various fair-share preferred factors (either from user
perspective or from resource usage). The plan (transfer paths) are
recorded back to the database indicating to {\it Link Managers} that
files are available for transfers. As proposed above, the link manager
supplies a particular link, using the back-end data mover. As soon as
the file appears available at the site and is planned to be
transferred via the link, the data mover instance is executed,
respecting the maximum simultaneously allowed transfers. The status of
the transfer is recorded back to the database, allowing users to see
the progress. The workflow is depicted in Fig. \ref{fig:architecture}.

\begin{figure}
  \begin{center}
    \includegraphics[width=3.5in]{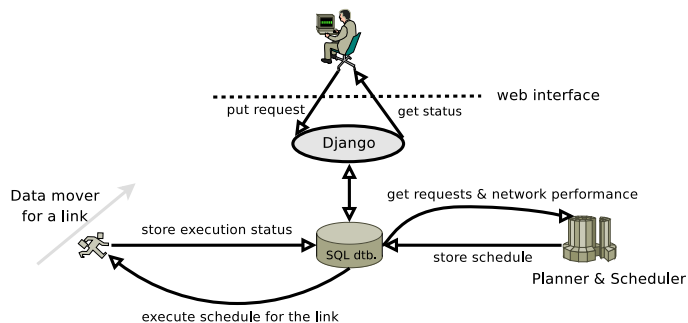}
    \caption{The scheme of a proposed architecture.}
    \label{fig:architecture}
  \end{center}
\end{figure}

\section{Conclusions}
In this paper we tackle the complex problem of efficient data
movements on the network within a distributed environment. The problem
itself arises from the real-life needs of the running nuclear physics
experiment STAR and its peta-scale requirements for data storage and
computational power as well. We presented the two stage constraint
model, coupling path planning and transfer scheduling phase for data
transfers to the single destination, with two alternative approaches
for planning transfer paths inspired by Simonis \cite{SimonisHCP}. We
proposed and implemented several search heuristics for both stages and
performed sets of experiments with realistic data input for evaluating
their applicability. Comparison of the results and trade-off between
the schedule of a constraint solver and a Peer-2-Peer simulator
indicates that it is promising to continue with the work, thus
bringing improvements over the current techniques to the
community. The execution of the schedule in a real environment and the
architecture of the system is proposed. In the nearest future we want
to concentrate on the integration of the solver with real data
transfer back-ends, consequently executing tests in the real
environment.

\ack{The investigations have been partially supported by the IRP AVOZ
  10480505, by the Grant Agency of the Czech Republic under Contract
  No. 202/07/0079 and 201/07/0205, by the grant LC07048 of the
  Ministry of Education of the Czech Republic and by the
  U.S. Department Of Energy.}

\section*{References}
\bibliography{zerola_chep_2009}
\bibliographystyle{iopart-num}

\end{document}